\RequirePackage[loading]{tracefnt}
\documentclass[12pt]{iopart}
\usepackage{citesort}
 \usepackage{graphicx}
\usepackage{color}
\graphicspath{ {Figures/} }

\begin{document}

\title{Direct loading of a large Yb MOT on the $^{1}S_{0} \rightarrow \, ^{3}P_{1}$ transition}

\author{A Guttridge$^1$, S A Hopkins$^1$, S L Kemp$^1$, D Boddy$^1$, R Freytag$^2$, M P A Jones$^1$, M R Tarbutt$^2$, E A Hinds$^2$ and S L Cornish$^1$}

\address{$^1$Joint Quantum Centre (JQC) Durham-Newcastle, Department of Physics, Durham University,
South Road, Durham, DH1 3LE, UK}
\address{$^2$The Blackett Laboratory, Prince Consort Rd., London, SW7 2BW, UK}

\ead{alexander.guttridge@durham.ac.uk}

\begin{abstract}
We report a robust technique for laser frequency stabilisation that enables the reproducible loading of in excess of 10$^{9}$ Yb atoms from a Zeeman slower directly into a magneto-optical trap (MOT) operating on the $^{1}S_{0} \rightarrow \, ^{3}P_{1}$ transition, without the need for a first stage MOT on the $^{1}S_{0} \rightarrow \, ^{1}P_{1}$ transition. We use a simple atomic beam apparatus to generate narrow fluorescence signals on both the 399~nm $^{1}S_{0} \rightarrow \, ^{1}P_{1}$ transition used for the Zeeman slower and the 556~nm $^{1}S_{0} \rightarrow \, ^{3}P_{1}$ transition. We present in detail the methods for obtaining spectra with a high signal-to-noise ratio and demonstrate error signals suitable for robust frequency stabilisation. Finally we demonstrate the stability and precision of our technique through sensitive measurements of the gravitational sag of the Yb MOT as a function of the intensity of the laser cooling beams, which are in good agreement with theory. These results will be important for efficient loading of the atoms into an optical dipole trap.
\end{abstract}

\section{Introduction}
\begin{figure}
\includegraphics{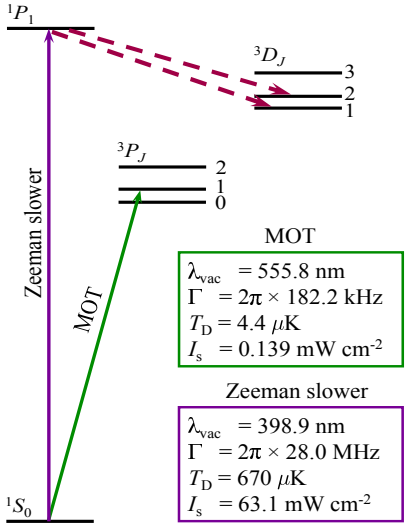}
\caption{\label{Elevel} Partial energy level diagram of ytterbium showing the dominant 398.9~nm singlet transition to the $^{1}P_{1}$ state and the narrow 555.8~nm intercombination transition to the $^{3}P_{1}$ state. The 555.8~nm transition is closed but the 389.9~nm transition is not, this is due to the weak decay of the $^{1}P_{1}$ state to the $^{3}D_{J}$ levels.}
\end{figure}

Ultracold ytterbium (Yb) atoms have found many interesting applications in modern AMO physics. Examples include studies of new quantum phases \cite{Liu2004,Taie2012}, atomic clocks \cite{Hoyt2005,Barber2006,Porsev2004,Hinkley2013}, tests of time-reversal symmetry \cite{Natarajan2005}, nuclear parity non-conservation \cite{DeMille1995} and quantum information processing \cite{Hayes2007,Reichenbach2007}. The large range of Yb isotopes, 5 bosonic and 2 fermionic, makes this atom ideal for the investigation of degenerate Bose-Bose and Bose-Fermi mixtures \cite{Fukuhara2009,Taie2010}. Notably, the fermionic isotope $^{173}$Yb is of great interest due to recent theoretical and experimental work leading to the observation of a unique orbital Feshbach resonance \cite{Zhang2015,Hoefer2015,Pagano2015}, allowing tunability of interactions in a system with SU($N$) symmetry \cite{Scazza2014}. Similarly, the scaling of the interspecies scattering length with reduced mass \cite{Munchow2011} makes Yb a strong candidate for the observation of novel Feshbach resonances in mixtures of alkali and alkaline-earth-like systems \cite{Brue2013,Kemp2016}. All these ultracold atom experiments start with the production of an Yb magneto-optical trap (MOT), which ideally should load quickly and reproducibly to a large number of atoms at the lowest achievable temperatures. This first stage is generally followed by evaporative cooling in an optical trap where large, reproducible MOTs are a prerequisite to reliable production of large Bose-Einstein condensates. It is therefore vitally important that the lasers used to cool and trap the Yb atoms have well stabilised frequencies. Ytterbium possesses two transitions suitable for laser cooling, the strong $^{1}S_{0} \rightarrow \, ^{1}P_{1}$ violet transition at 399~nm and the weakly allowed intercombination green transition $^{1}S_{0} \rightarrow \, ^{3}P_{1}$ at 556~nm. A diagram detailing the properties of these transitions is shown in \Fref{Elevel}.

The 399~nm transition is commonly used for Zeeman slowing of an Yb atomic beam as the large linewidth and short wavelength lead to a large maximum deceleration. Numerical simulations \cite{Hopkins2016} show that, for reproducible slowing, a frequency stability better than $\pm \, 2$~MHz is required for the Zeeman slower light, well below the natural linewidth, $2\pi \times 28$~MHz, of the transition. This transition is also used for absorption imaging as its high saturation intensity of 63~mW/cm$^{2}$ allows images to be recorded with good signal-to-noise in short times of order ten~$\mu$s. Accurate measurement of atom numbers from absorption imaging using this transition requires the absolute laser frequency to be stabilised to better than approximately $\pm \, 3$~MHz. Although the 399~nm transition is well suited to Zeeman slowing and imaging, it is not ideal for trapping and cooling in a MOT. The large linewidth gives a high Doppler temperature of 670~$\mu$K while decay from $^{1}P_{1}$ to the metastable $^{3}D_{J}$ levels causes a strong loss of atoms from the cooling cycle \cite{Honda1999}. Instead the closed, two-level transition at 556~nm is more suitable for the MOT. This transition is narrow, with a Doppler temperature of only 4.4~$\mu$K (although a temperature of 30~$\mu$K is more realistic \cite{Kemp2016}). However, the narrow linewidth results in a greatly reduced MOT capture velocity of 7~m$\, \mathrm{s}^{-1}$; this requires precise operation of the Zeeman slower. Furthermore, to achieve optimal loading of the MOT, the 556~nm laser must be simultaneously stabilised to below the natural linewidth, $2\pi \times 180$~kHz, of the cooling transition. The above requirements necessitate the production of Yb spectra with narrow linewidths and, ideally, a high signal-to-noise ratio (SNR).

In contrast to the alkalis, Yb has a low vapour pressure at room temperature \cite{Cottrell1973}. When heated to the temperatures required for significant absorption ($\geq420$ $^{\circ}$C) ytterbium reacts with glass, so spectroscopy of ytterbium atoms cannot be performed in conventional  vapour cells. Designs have been created that circumvent this problem \cite{Ishchenko2002,Jayakumar2015} but are necessarily bulky. Hollow-cathode lamps \cite{Kim2003,Loftus2000} are an alternative for absorption spectroscopy but these significantly broaden the features ($\geq$ 1 GHz), making sub-MHz frequency stabilisation difficult. Another alternative is stabilisation of lasers to a high finesse optical cavity \cite{Maruyama2003,Xiong2011}.

In this paper we demonstrate the production and measurement of large stable Yb MOTs using the green transition for the MOT and the violet transition for the Zeeman slower and for absorption imaging. We place considerable emphasis on the production of Yb fluorescence spectra with high SNR and the stabilisation of the lasers to those spectra, as this underpins the performance of our experiment. In sections \ref{sec:Setup} and \ref{sec:fluorescence} of this paper  we describe and demonstrate the production of fluorescence spectra from a simple Yb atomic beam apparatus. We include spectra of the $^{171}$Yb and $^{173}$Yb fermionic isotopes only, isolated by a polarisation technique \cite{Zinkstok2002,Long2014} from their normal overlap with lines from the remaining five bosonic isotopes. In section \ref{sec:locking} we discuss the frequency stabilisation techniques that we use to lock lasers to the 399~nm and 556~nm transitions. Finally, in section \ref{sec:cooling}, to demonstrate the stability of this locking, we present measurements of an Yb MOT operating on the \mbox{556 nm} transition. We show the direct loading of a MOT of $2.0 \times \, 10^{9}$ Yb atoms from a Zeeman slower without the need for a first stage operating on the 399~nm transition. We also present absorption images and measurements of the gravitational sag of the cold Yb atoms, which is highly sensitive to detuning and important for optimising the transfer to a dipole trap.



\section{Experimental Setup}\label{sec:Setup}

\begin{figure}
\includegraphics{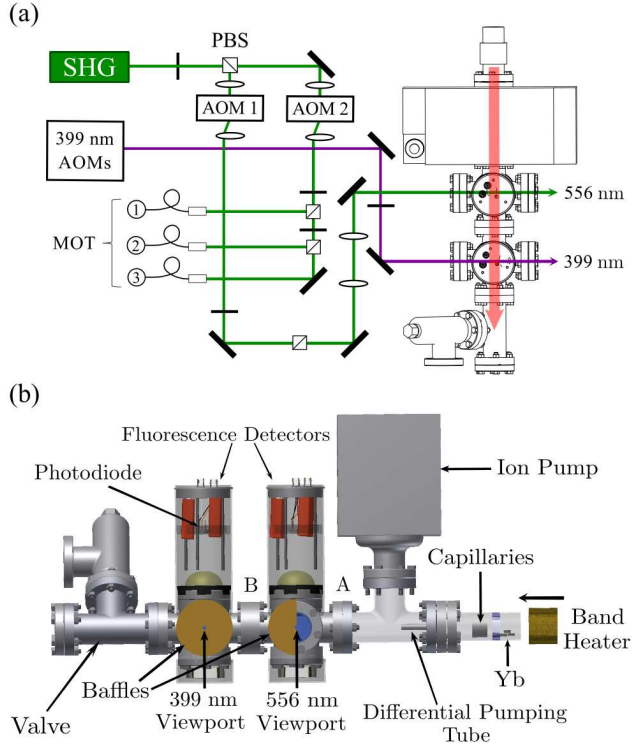}
\caption{\label{setup}(a) A schematic of the \mbox{556 nm} laser system with the Yb spectroscopy apparatus. The \mbox{556 nm} (green) light from the SHG is split into two paths. The majority of the light is fibre coupled into the three MOT fibres via AOM 2 which is used for control over the power and detuning of the light. The remainder of the light passes through the AOM 1 which applies a small modulation (dither) on the frequency before the light intersects the collimated atomic beam (red) at $90^{\circ}$. The \mbox{399 nm} light (purple) intersects the atomic beam after passing through an arrangement of three AOMs used to set the large Zeeman detuning and provide frequency control of the imaging light. The exact arrangement is not shown in the figure for clarity but is described in the text. (b) Overview of the Yb atomic beam spectroscopy apparatus. An Yb oven provides an atomic beam (right to left on figure) which is collimated by an array of capillary tubes and passes through a differential pumping tube and two further circular apertures denoted A and B (6 mm and 8 mm). Optical access for the laser beams is provided separately through the horizontal viewports of two six-way crosses, with the atomic fluorescence detected in the vertical direction by two photodiode assemblies.}
\end{figure}

An overview of the laser setup used for the fluorescence spectroscopy and MOT measurements is shown in \Fref{setup}(a). The 556 nm light is produced by frequency doubling a fibre laser operating at 1112 nm (Menlo Systems Orange One) to produce powers of up to 250 mW. The green light is then split into two paths by a polarising beamsplitter (PBS) cube. A pair of single-passed 200~MHz acousto-optic modulators (AOM) (Gooch and Housego, 46200-0.3-LTD) are used to set the required MOT detuning. The MOT light is delivered to the main experiment via three polarisation maintaining single-mode fibres. Intensity or detuning ramps are achieved using AOM~2 to control the power and frequency of the light. AOM~1 is used to apply a small modulation (dither) to the frequency of the light to allow frequency stabilisation to the atomic fluorescence (further details in \sref{sec:locking}). The light after AOM~1 is expanded by a telescope to 1.95 $\pm \, 0.05 $ mm, in order to increase the interaction volume of the laser light with the atomic beam.

The 399 nm light is produced by a commercial diode laser (Toptica DL Pro HP) which supplies up to 100 mW of single mode light. The majority of this is used for Zeeman slowing, with only a few mW being split off to the spectroscopy apparatus for both frequency stabilisation and to derive the light for absorption imaging. For the spectroscopy, prior to the intersection with the atomic beam, the light passes through two AOMs, each in a double pass configuration, which set the large detuning of 600 MHz required for Zeeman slowing \cite{Hopkins2016}. A third double passed AOM is used to give variable detuning of the imaging light. After the AOMs, the 399 nm light passes through the viewport of the spectroscopy apparatus with a measured 1/$\textrm{e}^{2}$ waist of 0.50 $\pm \, 0.01 $ mm before intersecting the atomic beam. Further details of our dual species CsYb atom-trapping experimental setup may be found in \cite{Kemp2016}.

\subsection{Atomic beam apparatus}

\Fref{setup}(b) illustrates the main components of the atomic beam spectroscopy apparatus, structurally similar to \cite{Long2014}, consisting of a steel vacuum chamber with two external assemblies for fluorescence detection and an external oven heater. The oven section is connected to the main section by a differential pumping tube of diameter 5 mm and length 50 mm.

Bright, collimated atomic beams may be produced \cite{Ross1995,Giordmaine1960} by using an array of parallel capillary tubes as the oven aperture, where the diameter of the tubes is smaller than the mean free path, $\Lambda$, of the effusing gas and the length of the tubes is commensurate with $\Lambda$. For a Maxwell-Boltzmann distribution, the mean free path is $\Lambda=k_{\mathrm{B}}T/\left(\sqrt{2}\pi d^{2} P\right)$ \cite{Ramsey1986}, where $P$ is the pressure of the gas, $T$ is the temperature of the gas and $\pi d^{2}$ is the scattering cross section. For the purposes of estimation we take $d$ to be the van der Waals diameter of an ytterbium atom, 560~pm \cite{Alvarez2013}. This yields a value for the mean free path of the order of 10~mm for our oven operating at $T=470 \, ^{\circ}$C. Our oven aperture consists of such an array of 55 capillary tubes each of internal diameter 0.58 mm and length 20 mm, positioned in front of a 5 mm diameter aperture in the oven wall. The capillary array can be removed to allow recharging of the Yb through the aperture. To produce a suitable Yb vapour pressure of order $10^{-3}$~mbar, the band heater surrounding the oven (Watlow $\mathrm{MB1J1JN2-X73}$) is raised to over 450~$^{\circ}$C. A horizontal, collimated Yb atomic beam then effuses from the capillaries and passes through two further apertures, the first of diameter 6~mm, the second of diameter 8~mm, positioned as shown in \Fref{setup}(b). 

\subsection{Fluorescence Detector}

\begin{figure}
\includegraphics{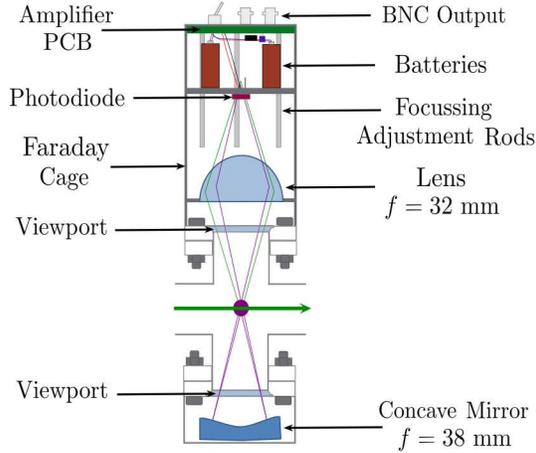}
\caption{\label{setuptwo}Detailed schematic of the fluorescence detection scheme. Fluorescence emitted upwards is collected by an aspheric lens of 50 mm diameter and 32 mm focal length, which is placed immediately above the upper viewport at $\sim$ 80~mm above the atoms. Fluorescence emitted downwards is captured by a retro-reflecting concave mirror of 50 mm diameter and 38 mm focal length, placed immediately below the lower viewport.}
\end{figure}

Each six-way cross is fitted with four viewports allowing a horizontal laser beam at either 556 or 399 nm to intersect the atomic beam at 90$^{\circ}$. The resulting resonance fluorescence is collected by custom built detectors based on designs presented in \mbox{Refs~\cite{Quessada2005,Boddy2014}}. The fluorescence is collected by a 50 mm diameter aspheric lens with a focal length of \mbox{32 mm} placed immediately above the upper viewport as shown in \Fref{setuptwo}. The lower viewport is fitted with a \mbox{50 mm} diameter, \mbox{38 mm} focal length retroreflecting mirror which serves to increase the amount of fluorescence detected by a factor of approximately three. This is more than the expected factor of two because fluorescence initially propagating downwards is reflected off the insides of the vacuum chamber and finds its way back via the lower mirror and upper collection lens to the photodiode. The two horizontal viewports are covered by light baffles, each with a small central aperture of 5~mm diameter; these block out unwanted extraneous light and also define a laser beam path that is orthogonal to the atomic beam. For the 556 nm laser it proved beneficial to use anti-reflection coated viewports for beam access to reduce the scattered light reaching the photodiode. The photodiode and its transimpedance amplifier are mounted together on a circular printed circuit board (PCB) which slides on three rods, thus allowing position adjustment to maximise the fluorescence signal.

Our custom-built fluorescence detector gives an excellent SNR without the expense of a photomultiplier and despite the low levels of fluorescence. A transimpedance amplifier, of very high gain yet low noise, is placed close to a standard large area photodiode (Centronics OSD50-T) inside a Faraday cage clamped over the vacuum viewport flange. The amplifier is powered by two 9 V batteries placed inside the cage and the amplified differential signal is carried out of the cage to the differential input of the lock-in amplifier via two BNC cables. Thus the entire circuit is protected against RF pickup and noise that might be otherwise injected from a power supply. The operational amplifier is a low input-noise device (Analog Devices AD795) \cite{Detector} with input pins protected by a guard ring to eliminate the effect of tiny currents that can flow in the PCB substrate. A `tee resistor network' \cite{BurrBrown1995} consisting of three resistors (of values 10~M$\Omega$, 10~M$\Omega$, 100~k$\Omega$) produces an effective transimpedance gain of up to $10^{9}$~V/A \cite{BurrBrown1995}, although we found that a gain in the range $(1 - 3) \times 10^{8}$~V/A was adequate for our purposes. This high gain allows us to run the atomic beam at relatively low density, thereby conserving atoms and extending the lifetime of the Yb source. At our oven operating temperature of 470~$^{\circ}$C, measured at the external band heater, we have so far found that 5 grams of Yb has lasted for over two years with usage of 10 hours/day. This scheme is widely applicable to other alkaline-earth-like elements with narrow laser cooling transitions. A similar apparatus is also used at Durham to frequency stabilise lasers to the $^{1}S_{0} \rightarrow \, ^{3}P_{1}$ cooling transition in strontium, where an additional, counter-propagating pump beam is used to generate linewidths narrower than the residual Doppler broadening \cite{Boddy2014}.



\section{Fluorescence spectroscopy of the atomic beam}\label{sec:fluorescence}

\begin{figure}
\includegraphics{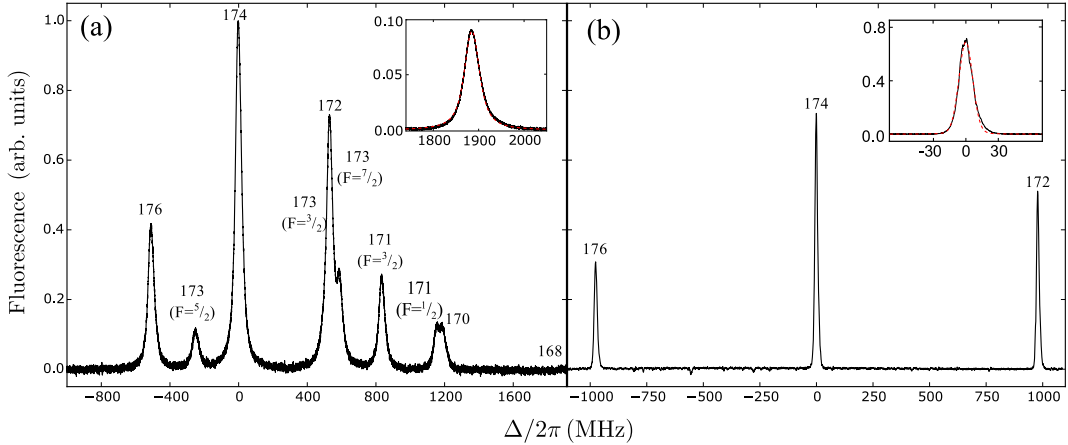}
\caption{\label{spectra} (a) Typical fluorescence signal obtained by scanning the 399 nm laser. The incident power is $P$ = 33 $\mu$W corresponding to $0.1 \, I_{\mathrm{sat}}$ and the heater temperature is \mbox{$T$ = 470 $^{\circ}$C}. The frequency axis is scaled using the literature values for the isotope shifts \cite{Das2005}. The inset shows a scan over the transition in $^{168}$Yb at $T$ = 540 $^{\circ}$C in black and a Voigt fit to the data in red. (b) Fluorescence signal obtained by scanning the 556 nm laser. The incident power is \mbox{$P$ = 460 $\mu$W} corresponding to $50 \, I_{\mathrm{sat}}$ and the temperature is $T$ = 470 $^{\circ}$C. In the 556~nm spectrum, the other isotopes lie outside the spectral range shown in the figure. The inset shows an enhanced view of the $^{174}$Yb resonance in black and a Voigt fit to the data in red.}
\end{figure}

Typical fluorescence spectra obtained with our system are shown in \Fref{spectra}. In \Fref{spectra}(a), the spectrum from the $^{1}P_{1}$ state at 399 nm is obtained with a modest laser power of 33 $\mu$W and an oven temperature of 470 $^{\circ}$C, demonstrating the excellent SNR of this system. Note that the spectra shown are the result of a 10-point moving average applied to the data by the oscilloscope. We see signals from all the isotopes of ytterbium with the exception of $^{168}$Yb, which has an extremely low natural abundance of 0.13\% \cite{Laeter2009}. However, in the inset of \Fref{spectra}(a), we clearly see $^{168}$Yb with a SNR greater than$100$ by increasing the oven temperature to \mbox{540 $^{\circ}$C}. The effect of Doppler broadening on the spectra at 399~nm creates a line shape described by a Voigt profile with FWHM of 40.0~$\pm \, 0.2$~MHz; this value is obtained from a least squares fit to the $^{168}$Yb peak \cite{Newville2014,Hughes2010}. The Gaussian contribution to the Voigt profile is dominated by 19.5~$\pm \, 0.1$~MHz of Doppler broadening, while the Lorentzian contribution is 30.0~$\pm \, 0.1$~MHz. 

A SNR greater than $300$ is also evident in spectrally narrow features of the $^{3}P_{1}$ state at 556~nm as shown in \Fref{spectra}(b). The inset in \Fref{spectra}(b) shows the $^{174}$Yb resonance fitted with a Voigt profile. The lineshape of the feature appears to be Gaussian, consistent with the narrow linewidth of the transition dominated by Doppler broadening. The largest component of homogeneous broadening is 1.3~MHz of power broadening due to the 460 $\mu$W of 556 nm light having an intensity of around $50 \, I/I_{\mathrm{sat}}$ when intersecting the atomic beam. Therefore, in the Voigt fit to the data we constrain the width of the Lorentzian component to be 1.3~MHz, resulting in a FWHM of 15.0~$\pm \, 0.2$~MHz. The Gaussian contribution has a Doppler width of 14.3~$\pm \, 0.2$~MHz due to the transverse spreading of the atomic beam. From this measurement of the Doppler broadening and using the mean speed of an effusive beam, $v_{\mathrm{mean}}=\sqrt{\frac{9\pi k_{\mathrm{B}} T}{8 m}} \simeq 350 \; \mathrm{m \, s}^{-1}$, we estimate the HWHM atom effusion angle to be 11~mrad. 

A Doppler shift also occurs due to any small departure of the crossing angle of the laser and the atomic beam from 90$^{\circ}$. The shift, $\delta\nu$, due to this effect is described by $\delta\nu=\left(v_{\mathrm{mean}}\cos\theta\right)/\lambda$, where $\lambda$ is the wavelength of the transition and $\theta$ is the crossing angle of the laser and the atomic beam. This equation shows that small departures from 90$^{\circ}$ result in an offset in the centre frequency of around 11 MHz per degree for the $^{1}S_{0} \rightarrow \, ^{3}P_{1}$ transition at $T=470 \, ^{\circ}\mathrm{C}$. A method for detection and correction of this offset is described in \Sref{sec:locking}.

\begin{figure}
\includegraphics{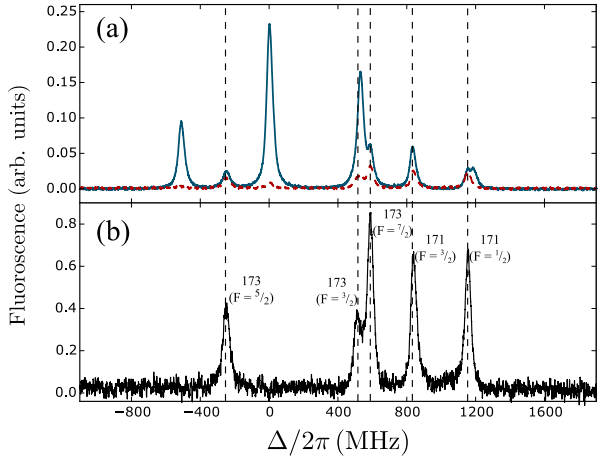}
\caption{\label{fermion}(a) The fluorescence spectrum of the $^{1}P_{1}$ state for vertically (horizontally) polarised light is shown by the red (blue) line for a power P = 265 $\mu$W and temperature T = 470 $^{\circ}$C. (b) Scaled difference of the two signals showing only the fermionic spectrum. The dashed vertical lines show the detunings measured by Das et al \cite{Das2005}. }
\end{figure}

The five bosonic isotopes do not exhibit hyperfine structure as they have a nuclear spin of $I=0$. Hence, the relative peak heights of transitions in the bosons are directly proportional to their natural abundances, this is clearly exhibited in \Fref{spectra}(a) where the $^{174}$Yb peak is 2.5 times higher than the $^{176}$Yb peak, in good agreement with their respective natural abundances of 31.8\% and 12.7\% \cite{Laeter2009}. On the other hand, the fermionic isotopes $^{171}$Yb and $^{173}$Yb possess hyperfine structure due to their non-zero nuclear spin of $I=1/2$ and $5/2$ respectively.


\Fref{fermion} illustrates the production of a purely fermionic spectrum. The detected fluorescence spectra for two orthogonal cases of linear polarisation of the 399 nm laser for P = 265 $\mu$W are shown in \Fref{fermion}(a), where horizontal polarisation is shown in blue and vertical polarisation in red. The suppression of the fluorescence for the bosonic isotopes is due to their lack of ground state structure and can be exploited to obtain a purely fermionic spectrum \cite{Zinkstok2002,Long2014}, allowing precise measurement of the hyperfine splittings and isotope shifts. Due to the large collection angle of the fluorescence detector we do not see a perfect extinction of the boson signal. However, by subtracting the horizontally polarised signal from the vertically polarised signal (multiplied by a suitable factor of about 0.033 to match the heights of the bosonic peaks), we obtain the fermion spectrum shown in \Fref{fermion}(b). Here we resolve all the hyperfine components of $^{173}$Yb and $^{171}$Yb, including the $F = 3/2$ and $7/2$ transitions of $^{173}$Yb which are typically masked by $^{172}$Yb. The high SNR of the raw signals mean that even after this subtraction process, the resulting SNR remains reasonable.

\section{Frequency stabilisation}\label{sec:locking}

The spectroscopic techniques detailed above are used to stabilise the frequency of our 399~nm and 556~nm lasers to any isotope by locking to the atomic beam spectrum.  For the 399 nm laser we use phase sensitive detection by modulation of the laser current at a frequency of 3.7~kHz, with a depth of $\pm \, 0.5$~MHz on the laser frequency. The resulting fluorescence signal is demodulated by a lock-in amplifier to generate the dispersive error signal.

\begin{figure}
\includegraphics[width=0.5\textwidth]{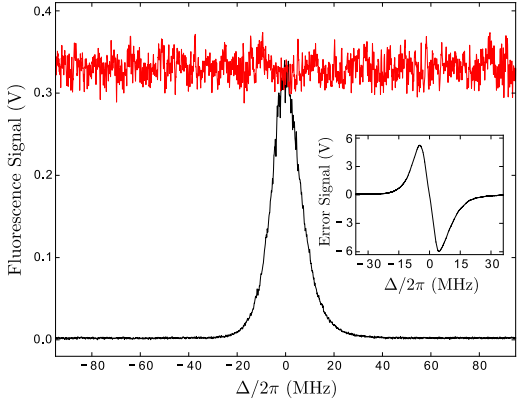}
\caption{\label{error} Atomic fluorescence versus detuning from the $^{1}S_{0} \rightarrow \, ^{3}P_{1}$ transition . The black trace shows the $^{174}$Yb fluorescence peak when scanned over the $^{1}S_{0} \rightarrow \, ^{3}P_{1}$ transition  and the red trace shows the fluorescence when the laser is locked. The inset shows the error signal generated by the frequency modulation spectroscopy, again with excellent SNR.}
\end{figure}

The challenging task of locking to the 556~nm transition is typically accomplished by utilising a combination of an optical cavity and atomic spectroscopy to achieve the required frequency stability \cite{Maruyama2003,Xiong2011}. Here, we demonstrate that simple phase sensitive detection of the fluorescence signal from the atomic beam apparatus is sufficient to load a large MOT of Yb atoms. For phase sensitive detection of the 556~nm beam, we use an AOM to apply a dither at a frequency of 3~kHz to the AOM shift centred at 217~MHz that results in optical frequency excursions of depth $\pm \, 2$~MHz. The detected fluorescence is demodulated by a lock-in amplifier yielding a dispersive error signal. \Fref{error} shows the fluorescence peak of $^{174}$Yb scanned over the $^{1}S_{0} \rightarrow \, ^{3}P_{1}$ transition (black) and the detected fluorescence when the laser is locked to the $^{1}S_{0} \rightarrow \, ^{3}P_{1}$ transition (red). The associated dispersive error signal is displayed in the inset and has a central slope of 1.22~V/MHz measured at the output of the lock-in amplifier. The level of noise on the fluorescence signal when the laser is locked suggests the short term fluctuations are below 1~MHz.

\begin{figure}
\includegraphics[width=0.5\textwidth]{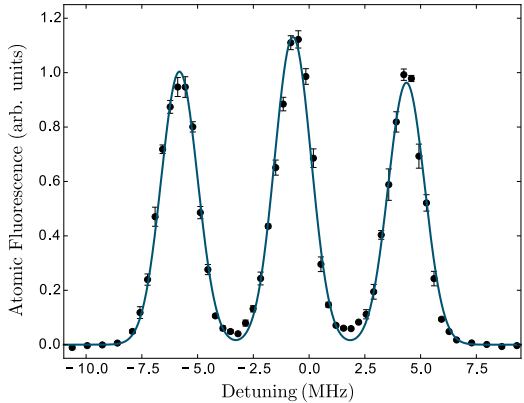}
\caption{\label{insensitive} Fluorescence of ultracold $^{174}$Yb atoms versus `detuning' of 556 nm MOT beams with an applied magnetic field of 2.4 G. Here, the detuning is simply the frequency difference between the MOT and spectroscopy AOMs when the laser is locked on to the zero crossing of the fluorescence error signal. The blue line is a fit to the sum of three Gaussians and is used to extract the centre of the $m'_{J} = 0$ peak.}
\end{figure}

In order to measure any systematic offset in the lock to the $^{1}S_{0} \rightarrow \, ^{3}P_{1}$ transition, we exploit the magnetically insensitive transition $m_{J} = 0 \rightarrow m'_{J} = 0$ in $^{174}$Yb as follows. We first load a MOT of $^{174}$Yb atoms at a total MOT beam intensity of $I_{\mathrm{total}}=270 \, I_{\mathrm{sat}}$, where $I_{\mathrm{total}}$ is the sum of the intensities of all six beams. We then turn off the magnetic field gradient and the MOT light and immediately apply a 2.4~G bias field to produce a Zeeman shift for the magnetically sensitive states. The MOT beams are then pulsed back on at an intensity of $I_{\mathrm{total}}=15 \, I_{\mathrm{sat}}$ and the fluorescence from the cold atoms is detected. A plot of the detected atomic fluorescence against the detuning of the pulsed MOT light from the locked laser is shown in \Fref{insensitive}. The figure shows the three expected peaks for the $m'_{J} = -1$, $m'_{J} = 0$ and $m'_{J} = 1$ states, with the centre frequency of the $m'_{J} = 0$ peak found to be offset by $-0.72 \pm \, 0.06$ MHz prior to calibration. The most likely cause of the offset is a small deviation of the crossing angle between the laser and the atomic beam of 0.070 $\pm \, 0.006^{\circ}$ from 90$^{\circ}$. This offset can then be corrected to zero by compensation with AOM 1.

\section{Laser cooling of Yb}\label{sec:cooling}

\begin{figure}
\includegraphics[width=0.5\textwidth]{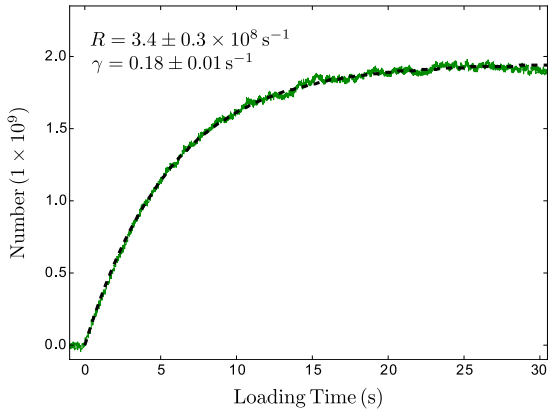}
\caption{\label{load} Loading curve of the Yb MOT operating on the 556 nm~transition for a total beam intensity of $I_{\mathrm{total}}=270 \, I_{\mathrm{sat}}$, an axial magnetic field gradient of 3.4~G/cm and a detuning of 4.6~MHz. The MOT loads a steady state number of $2.0 \times \, 10^{9}$ atoms. The dashed lines shows a least squares fit to \Eref{eq:loading}. }
\end{figure}

The efficiency of the loading of Yb MOTs is strongly dependent on the performance of the frequency stabilisation of the narrow 556~nm transition. In \Fref{load} we demonstrate the production of a large Yb MOT with a steady state atom number of $2.0 \times \, 10^{9}$ atoms. We achieve this number using 556~nm light with a total beam intensity of $I_{\mathrm{total}}=270 \, I_{\mathrm{sat}}$ and a detuning of 4.6~MHz. The axial magnetic field gradient is 3.4~G/cm. We extract the atom number from fluorescence measurements of the MOT, this is calibrated by absorption imaging on the 399~nm transition and agrees well with the number calculated from the fluorescence directly.

The loading of atoms in a MOT can be simply modelled by the equation
\begin{equation}\label{eq:loading}
N=\frac{R}{\gamma}\left(1-\exp \left(-\gamma t \right)\right),
\end{equation}
where $N$ is the number of atoms in the MOT, $t$ is the loading time, $\gamma$ is the loss rate and $R$ is the loading rate. A least squares fit of the equation to the data is shown by the dashed line in \Fref{load}, from which we find a loading rate of $R=3.4 \times \, 10^{8}$~s$^{-1}$. This efficient loading of the MOT is accomplished without the need for frequency sidebands on the MOT light, a technique commonly employed \cite{Peng-Yii2009} to increase the low capture velocity of 556~nm transition. This result shows that the 399~nm Zeeman laser is frequency stabilised below the $\pm \, 2$~MHz required for optimum operation \cite{Hopkins2016}.

\begin{figure}
\includegraphics[width=0.5\textwidth]{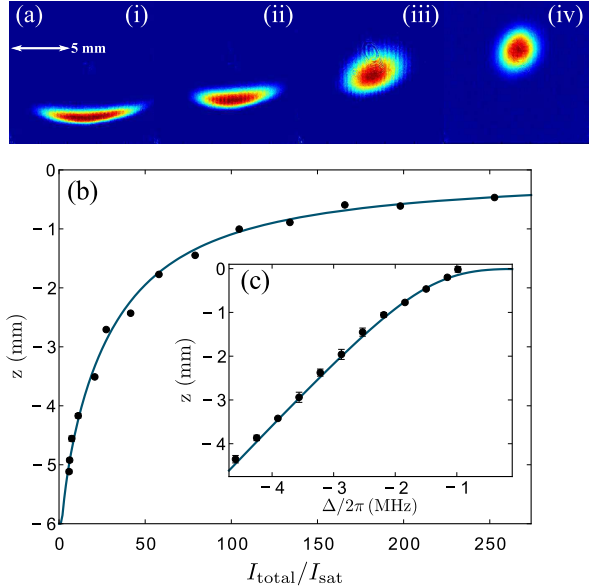}
\caption{\label{sag}(a) Absorption images of the MOT for different total intensities of MOT light (i) $1.7 \, I_{\mathrm{sat}}$ (ii) $7.1 \, I_{\mathrm{sat}}$ (iii) $38 \, I_{\mathrm{sat}}$ (iv) $130 \, I_{\mathrm{sat}}$. (b) Vertical sag due to gravity as a function of total intensity of the MOT light. The circles are experimental data and the blue line is a fit of \Eref{MOT} with detuning $\Delta = -4.6$~MHz and axial gradient $\frac{dB_{\hat{z}}}{dz} =3.4$~G/cm. (c) Vertical sag as a function of the detuning of the MOT light. The circles are experimental data and the blue line is a fit of \Eref{MOT} with total intensity $I_{\mathrm{total}} = 2.1 \, I_{\mathrm{sat}}$ and axial gradient $\frac{dB_{\hat{z}}}{dz} =3.4$~G/cm.}
\end{figure}

To further demonstrate that the lasers are stably and reliably locked with our Yb atomic beam apparatus, we include in \Fref{sag} a series of absorption images and measurements of an Yb MOT sagging under gravity with low intensities of MOT light. Under these low intensity conditions, the vertical MOT position is extremely sensitive to the MOT detuning \cite{Ludlow2008,Boddy2014}. To produce the images in \Fref{sag}(a) we directly load a MOT of $^{174}$Yb atoms using the 556~nm transition at an intensity of $I_{\mathrm{total}}=270 \, I_{\mathrm{sat}}$. After loading the MOT, we ramp down the intensity of the MOT light before taking an absorption image of the lowered atoms using the strong $^{1}S_{0} \rightarrow \, ^{1}P_{1}$ transition.

In \Fref{sag}(b) we measure the depth of the sag as a function of intensity and we see that the MOT drops a distance of up to 5 mm, which is greater than its initial size. This position shift is an important parameter when utilising intensity and detuning ramps to reach lower temperatures \cite{Kemp2016} before loading the atoms into an optical dipole trap. The equilibrium $z$ position may be determined by equating the force due to gravity with the net scattering force due to the two vertical MOT beams. This yields
\begin{equation}
\fl m\mathrm{g}=\frac{\Gamma \hbar k}{2} \left[\frac{I/I_{\mathrm{sat}}}{1+6I/I_{\mathrm{sat}}+\frac{4}{\Gamma^{2}}\left(\Delta - \frac{\mu}{\hbar} \frac{dB_{\hat{z}}}{dz}z\right)^{2}}-\frac{I/I_{\mathrm{sat}}}{1+6I/I_{\mathrm{sat}}+\frac{4}{\Gamma^{2}}\left(\Delta + \frac{\mu}{\hbar} \frac{dB_{\hat{z}}}{dz}z\right)^{2}}\right],\label{MOT}
\end{equation} 
where $\Delta$ is the detuning of the light, $\mathrm{g}$ is the standard acceleration due to gravity, $\frac{dB_{\hat{z}}}{dz}$ is the magnetic field gradient in the $\hat{z}$ direction, $\mu = g_{J}\mu_{\mathrm{B}}$ is the effective magnetic moment, $g_{J}$ is the Land\'e g-factor of the $^{3}P_{1}$ state and $\mu_{\mathrm{B}}$ is the Bohr magneton. The blue line in \Fref{sag}(b) is a fit of \Eref{MOT} to the experimental data (circles), with fit parameters: MOT detuning $\Delta=-4.60 \pm 0.04$~MHz and axial gradient $\frac{dB_{\hat{z}}}{dz} =3.40 \pm 0.05$~G/cm. In \Fref{sag}(c) we measure the sag as a function of detuning $\Delta$. The blue line is again a fit to the experimental data (circles) using \Eref{MOT}, in this case with fit parameters of: total MOT intensity $I_{\mathrm{total}} = 2.10 \pm 0.05 \, I_{\mathrm{sat}}$ and axial gradient $\frac{dB_{\hat{z}}}{dz} =3.40 \pm 0.05$~G/cm. The excellent agreement between the experimental and fitted data in \Fref{sag}(c) over a range of detunings of 4 MHz shows that our detuning is stable to below the power-broadened linewidth of the transition and is highly reproducible during an experimental run of several hours duration. This reproducibility, combined with our understanding of the gravitational sag, are essential prerequisites to loading the atoms into an optical dipole trap for subsequent cooling to quantum degeneracy.



\section{Conclusion}\label{sec:conclusion}
We have demonstrated a simple approach to laser frequency stabilisation that enables the reproducible loading of a large number of Yb atoms directly into a MOT operating on the 556~nm $^{1}S_{0} \rightarrow \, ^{3}P_{1}$ transition. In our apparatus, an oven produces a well collimated atomic beam which is probed by resonant laser light, crossing at 90$^{\circ}$ to reduce Doppler broadening. The fluorescence from the atoms is detected using a high numerical aperture lens and a high-gain photodiode circuit. We presented fluorescence spectra obtained with this apparatus for both Yb laser cooling transitions (at 399 nm and  556 nm) and showed that they are suitable for generating high SNR dispersive error signals for laser locking without the need for saturated absorption techniques \cite{Jayakumar2015,Loftus2000}. By switching the laser polarisation we were also able to derive a spectrum for the two fermionic isotopes only. We also show that the laser lock may be calibrated to better than 1 MHz using Yb MOT fluorescence measurements of the $m_{J} = 0 \rightarrow m'_{J} = 0$ magnetically insensitive transition of the $^{3}P_{1}$ state. The performance of the locking is demonstrated by the loading of a large ($>10^{9}$ atoms) Yb MOT without the use of frequency sidebands on the MOT light or an optical cavity in the locking scheme. The stability and reproducibility of the lock to the narrow $^{1}S_{0} \rightarrow \, ^{3}P_{1}$ transition is demonstrated by measurements of an Yb MOT sagging under gravity at low MOT light intensities. We have used this system to lock both the 399 nm and 556 nm lasers for over 2 years, during which time the spectroscopy has been highly reproducible. The approach is also applicable to other alkaline earth and lanthanide elements with narrow laser cooling transitions \cite{Boddy2014,De2009,Sukachev2010,Guest2007,Maier2014,Frisch2012}. We expect our work will be of interest to other groups starting to construct experiments to study ultracold quantum gases of Yb, where robust and reproducible laser frequency stabilisation is an essential prerequisite to cooling to quantum degeneracy.

\ack

We acknowledge support from the UK Engineering and Physical Sciences Research Council (grant number EP/I012044/1) and from the Royal Society. The data presented in this paper are available from http://dx.doi.org/10.15128/9s1616164.
\section*{References}

\bibliography{YbSpecV3}

\end{document}